# Uni-Macro-FRPN: Full-Resolution and Cross-Scale Learning for Polymers

Jintao Wu[1,2], Yiran Shan[1,2], Rui Zhang[1,2]*

[1]South China Advanced Institute for Soft Matter Science and Technology, School of Emergent Soft Matter, South China University of Technology, Guangzhou 510640, China.

[2] Guangdong Provincial Key Laboratory of Functional and Intelligent Hybrid Materials and Devices, South China University of Technology, Guangzhou 510640, China.

**ABSTRACT:** Polymer properties emerge from interactions across scales, yet existing polymer models typically preserve either detailed monomer chemistry without an explicit polymer graph or polymer connectivity with simplified monomer representations, due to computational constraints, as polymers typically contain tens of thousands of atoms. We present Uni-Macro-FRPN (FRPN), a Full-Resolution Polymer Network that retains both detailed atom-level and monomer-level features and explicit polymer structure information within a unified framework. Two Transformers jointly learn atom-informed monomer semantics, sequence order, and chain topology from BigSMILES-derived representations. On the Block Copolymer Database (BCDB) lamellar-versus-non-lamellar classification task, FRPN achieves 86.4% accuracy and 90.6% ROC–AUC, establishing state-of-the-art performance. Ablation results indicate that the gain is not explained solely by increased parameter count. On a linear homopolymer benchmark, monomer-centric learning remains competitive, highlighting a boundary case where polymer-scale organization is simple. To test generalization beyond linear polymers, we further construct an all-atom molecular-dynamics benchmark of 1640 datapoints spanning diverse monomer chemistries, sequence orderings, chain topologies, and physical properties. FRPN achieves the strongest overall performance on this topology-rich benchmark, with diagnostics supporting the benefit of jointly modeling monomer chemistry and polymer structure. Taken together, FRPN provides a practical route for moving polymer representation learning beyond monomer-centric representations. The leading performance of FRPN also suggests a promising direction for polymer informatics: future polymer prediction models should treat polymers not only as collections of monomer descriptors, but as complete multiscale chemical and topological objects.

## INTRODUCTION

Machine-learning and deep-learning methods have become widely used tools in scientific research, enabling rapid progress in both property prediction and inverse design for small molecules and polymers[1-5]. Polymer property prediction, however, presents a distinct representation problem because polymer properties are governed by features across coupled scales, from local chemistry to polymer sequence[6, 7], chain conformation[8], and interchain packing or entanglement[9]. This multiscale nature makes polymer modeling more demanding and requires representations that preserve detailed monomer-level information while also encoding chain-level organization.

Most existing deep-learning models for polymer property prediction avoid explicit full-chain modeling and instead represent polymers using repeat-unit structures[10-15]. This strategy is computationally efficient, as polymer chains are often long and highly repetitive, and repeat-unit representations can preserve essential chemical information. However, such representations generally omit or oversimplify the chain-level features discussed above.

Many studies have introduced higher-order polymer information through several routes. Descriptor-based workflows introduce chain variables as explicit input features[16, 17], which provides property-relevant metadata but does not specify how chain-level factors interact with local repeat-unit chemistry. The BigSMILES[18] notation incorporates polymeric fragments and connectivity into a string representation, while molecular-ensemble graph models, such as wD-MPNN[19], represent polymers as possible monomer molecule ensembles. However, these representations mainly describe polymer connectivity or statistical composition at an ensemble level, while explicit chain topology and monomer ordering are not directly encoded. Periodicity-aware methods further encode the repeating nature of polymers through periodic polymer graphs or periodicity priors[20, 21]. Nevertheless, these methods mainly exploit local periodic fragments and are not well suited to polymers with irregular sequences or complex backbones. More broadly, these methods remain centered on monomeric or oligomeric representations, while explicit polymer structures are largely omitted. This limitation is reflected in Uni-Macro[22], which observed diminishing marginal returns from atomic descriptors and pairwise structural biases under a monomer-centric deep learning model, thereby calling for frameworks of explicit higher-order polymer features.

Coarse-grained and sequence-based models represent polymers as bead or monomer sequences, allowing sequence distribution and backbone topology to enter the prediction model[23-25]. These studies demonstrate the value of backbone-aware representations. However, they simplify monomer chemistry into bead types or fixed feature vectors, and rely on relatively simple neural architectures. More recent structure-aware models have extended this line of work to graph neural networks (GNNs). Kimmig et al. represent polymer samples as monomer-node graphs, but monomer chemistry is still supplied through fixed fingerprints or embeddings[26]. Han et al. instead learn monomer representations from atomic graphs, but the polymer backbone is encoded as a compressed fragment-level graph[27]. More importantly, these GNN-based models rely on local message passing for backbone reasoning, making long-range communication susceptible to information bottlenecks or over-squashing across graph paths[28-31]. This limitation is nontrivial for polymers, whose properties may depend on nonlocal interactions and correlations along chains[32-34].

Developed for proteins, hierarchical representations provide useful references for polymer modeling. PepMNet learns atomic peptide graphs, pools features into amino-acid nodes, and applies a second graph network.[35] Zhang et al. segment proteins by secondary structure, quantize the fragments with a VQ-VAE, and model the resulting structural-token sequence.[36]

Motivated by these limitations and related works, we present Uni-Macro-FRPN, a full-resolution polymer representation framework that learns across atom, monomer and polymer-chain scales. FRPN couples a monomer-level Transformer[37] with a polymer-scale Transformer, allowing local chemical representations to be contextualized by explicit backbone organization rather than treated as isolated repeat-unit descriptors. We evaluate FRPN on the BCDB benchmark[38] where incorporating polymer-backbone information through the BigSMILES notation substantially improves predictive performance over strong baselines. We then test FRPN on a linear homopolymer benchmark, showing that monomer-centric representations remain competitive when the polymer backbone is structurally simple, thereby clarifying the regime in which full-resolution chain modeling is most beneficial. To address the broader scarcity of polymer datasets with explicit structural diversity, we further construct an all-atom molecular dynamics benchmark spanning diverse monomer chemistries, chain architectures and physical properties. On this benchmark, matched-pair analyses suggest that single-branch models can exploit cross-source proxy signals, while the follow-up residual analyses support a benefit from joint modeling beyond the tested prediction-level combinations. Together, these results suggest that polymer property prediction benefits not only from richer monomer descriptors, but from making the polymer chain itself an explicit object of representation. FRPN therefore marks a shift from monomer-centric learning toward full-resolution cross-scale modeling, in which local chemistry and macromolecular organization are learned within a unified architecture.

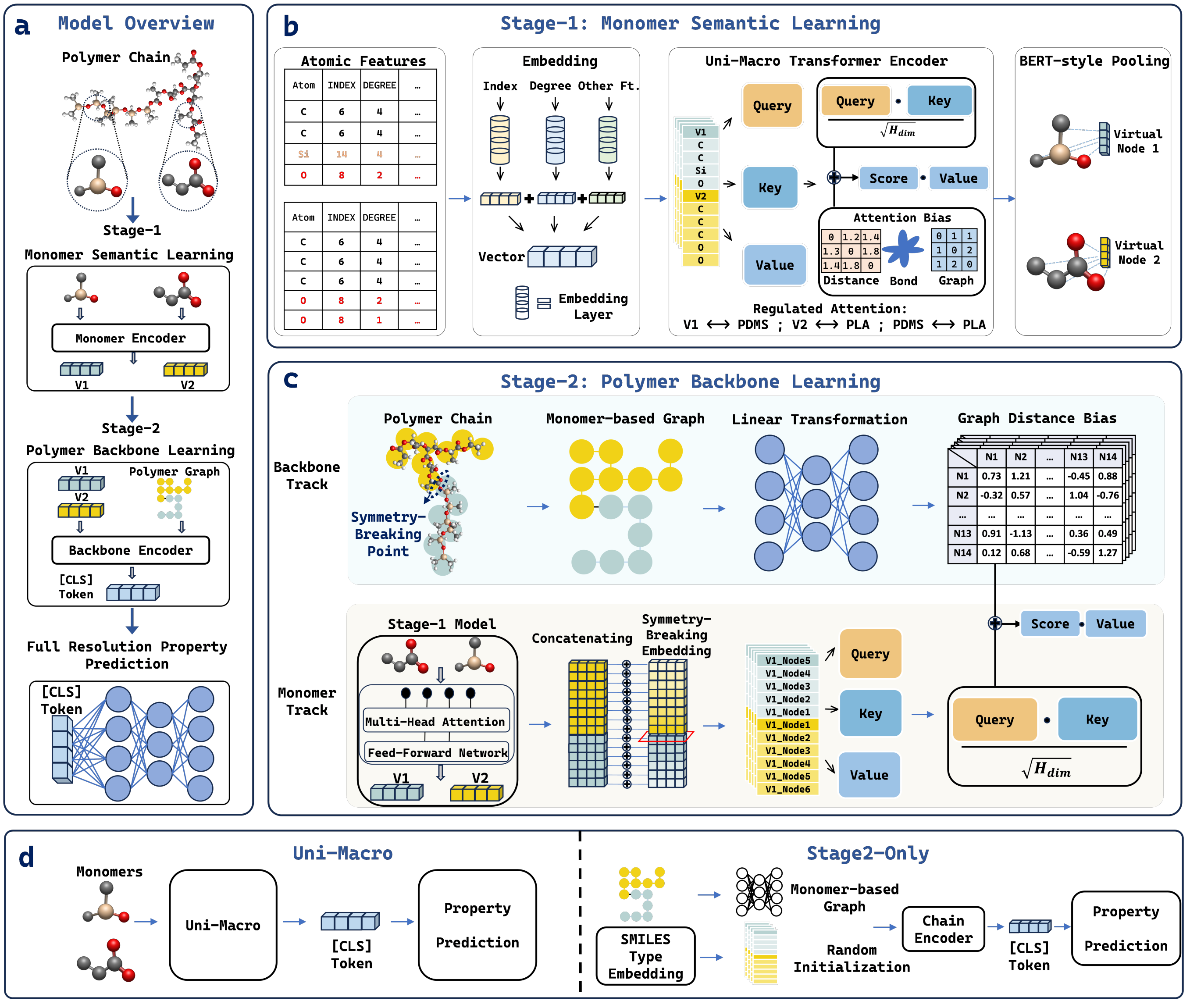

**Figure 1. Overview of FRPN architecture.** (a) Overall workflow connecting atomistic monomer encoding, polymer-chain representation, and property prediction. (b) Stage-1 encodes each monomer using a Uni-Macro-based Transformer and produces a virtual-node embedding of its chemical and structural features. (c) Stage-2 places these embeddings on the monomer-instance graph and uses graph-distance attention biases and symmetry-breaking modulation to learn polymer-scale context. The output is pooled through a global token and passed to a prediction head. (d) Single-branch controls: Uni-Macro retains monomer learning, whereas Stage2-Only uses trainable monomer-identity embeddings for chain-graph reasoning.

## RESULTS AND DISCUSSION

**Model Framework.** As illustrated in Figure 1a, FRPN connects atomistic monomer encoding with explicit polymer-chain modeling in a two-stage architecture. Stage-1 embeds atomic identities and within-monomer structural features. Stage-2 uses these learned monomer representations for structure-aware reasoning over the polymer backbone.

The workflow of stage-1 is illustrated in **Figure 1**b. FRPN adopts Uni-Macro[22] as the monomer encoder. Each polymer repeat unit is first represented as an atomistic graph with rich atom-level descriptors, together with monomer structural information including bond adjacencies, distance matrix, and graph distances. These signals are incorporated into the attention mechanism so that the encoder learns monomer representations not only from atomic features, but also from monomer structural context. Stage-1 then compresses each monomer into a summary representation. This virtual-node embedding serves as the semantic carrier of the monomer and forms the interface between the atomistic and polymer-level stages.

The Stage-2 workflow is illustrated in Figure 1c. The second stage treats the polymer as a graph defined over monomer units. Starting from the polymer architecture, we construct a monomer-resolution chain graph in which each node corresponds to one monomer instance, and edges encode the connectivity imposed by the polymer backbone. The graph distance is then transformed into an additive attention bias, allowing the model to explicitly encode the polymer structural context. The virtual-node embeddings produced by Stage-1 are aligned with these monomer nodes and concatenated into the Stage-2 input sequence, so that attention is computed jointly over monomer semantics and chain connectivity. In addition, for highly repetitive polymers with simple topologies, a symmetry-breaking signal is introduced to distinguish identical repeated units in the initial sequence. The output sequence of Stage-2 is then aggregated into a global [CLS] representation, which is passed to a multilayer prediction head for downstream property prediction. This architectural design enables the model to learn polymer representations across three nested structural levels—atom, monomer, and polymer—thereby providing a full-resolution framework for polymer property prediction.

**Table 1. Five-fold split statistics for the BCDB benchmark.**

| Fold | Train Size | Validation Size | Validation Positives | Validation Negatives | Train Unique Pairs | Val Unique Pairs |
|---|---|---|---|---|---|---|
| 0 | 4331 | 1038 | 660 | 378 | 35 | 9 |
| 1 | 4380 | 989 | 496 | 493 | 37 | 7 |
| 2 | 4378 | 991 | 349 | 642 | 39 | 5 |
| 3 | 4213 | 1156 | 786 | 370 | 28 | 16 |
| 4 | 4174 | 1195 | 461 | 734 | 37 | 7 |

**BCDB benchmark and training protocol.** We choose BCDB[38] as the benchmark dataset. BCDB provides polymer-level information including density, number average molecular weight (Mn), as well as the BigSMILES[18] notation of polymers, which allows us to recover polymer topology while retaining accurate monomer representations. We formulate BCDB as a relatively balanced binary classification of lamellar versus non-lamellar records, using the class assignments shown in Figure S1. We employ five-fold cross-validation for evaluation where in each split, one fold is used for validation and the remaining four folds are used for training. We also design the splits to maintain sufficient training and validation samples while preserving diversity in chemical pairs. To avoid data leakage, training and validation folds are constructed in the way that chemical pairs appearing in the training set do not appear in the validation set. The fold-wise details are summarized in **Table**

1. Details of chemical-pair distributions are summarized in Table S1, and the corresponding SMILES and BigSMILES representations are provided in Table S2.

In addition to three strong baseline models PerioGT[21], TransPolymer[12], Structure-aware GCN (SA-GCN)[26], we include two FRPN ablation variants as shown in **Figure 1**d: Uni-Macro and Stage2-Only. Uni-Macro retains monomer semantic learning but removes polymer backbone reasoning. Stage2-Only preserves polymer backbone reasoning, while monomer nodes are initialized from SMILES-ID embeddings instead of learned molecular representations. All models use identical data splits, with model-specific training settings. Training hyperparameters across benchmarks involved in this work are provided in Table S3, and architectural settings are provided in Table S4.

**Table 2. BCDB classification performance of FRPN and baseline models, with best values in bold and standard deviations across five folds.**

| Model | AUC | ACC | BACC | F1-Score | MCC |
|---|---|---|---|---|---|
| PerioGT | 0.619±0.041 | 0.665±0.061 | 0.624±0.027 | 0.644±0.147 | 0.288±0.051 |
| TransPolymer | 0.614±0.095 | 0.654±0.046 | 0.614±0.078 | 0.638±0.171 | 0.242±0.151 |
| SA-GCN | 0.820±0.107 | 0.762±0.109 | 0.728±0.154 | 0.782±0.080 | 0.452±0.302 |
| Uni-Macro | 0.788±0.044 | 0.775±0.053 | 0.754±0.033 | 0.772±0.097 | 0.523±0.079 |
| Stage2-Only | 0.741±0.141 | 0.737±0.120 | 0.724±0.111 | 0.695±0.161 | 0.461±0.210 |
| FRPN | **0.906±0.029** | **0.864±0.054** | **0.853±0.073** | **0.848±0.108** | **0.717±0.120** |

**Performance evaluation on BCDB benchmark.** The performance of FRPN and the baseline models is summarized in Table 2. **Table S5 lists the inputs supplied to each model on all three benchmarks.** FRPN ranks first among the six evaluated models on all five metrics, including AUC[39, 40], ACC, BACC[41], F1-score[42], and MCC[43]. More detailed metrics, including AP, precision, sensitivity, and specificity, are reported in Table S6. This consistent advantage suggests that the joint modeling of local chemistry and polymer backbone structure provides a clear benefit over the tested single-branch representations. In addition, the performance improvement of FRPN over GNN-based chain-reasoning models highlights the importance of long-range backbone communication, where Transformer attention enables direct interactions between distant chain segments instead of relying on iterative local message passing.

To examine performance on unseen monomers, we stratified OOF predictions by training-fold monomer exposure. The 44 held-out pairs comprised 28 with two seen monomers, 13 with one unseen monomer, and three with two unseen monomers (4899, 451, and 19 samples). FRPN ranked first among six models on all five metrics in every subset, with accuracies of 86.28%, 84.04%, and 94.74% and ROC–AUC values of 0.8967, 0.8682, and 0.9318, respectively. FRPN maintains strong BCDB performance for held-out pairs containing unseen monomers. Detailed subgroup metrics and nearest-training-monomer ECFP4 similarities are reported in Tables S7 and S8, respectively.

**Decision-level analysis from Out-of-Fold predictions**. Beyond the aggregate scores, we examine how these improvements are reflected in the prediction behavior of each model. As shown in **Figure 2**a, FRPN achieves consistently strong ROC behavior across the five validation folds. It obtains the highest AUC in Folds 0, 1, 2 and 3, and remains close to the best result in Fold 4. In contrast, several baseline models exhibit larger fold-dependent fluctuations.

**Figure 2**b evaluates each model on its model-specific disagreement set, comprising records on which the other five models predict different classes. FRPN has the highest accuracy in four of five folds and a five-fold mean of $0.830 \pm 0.086$ on these subsets. The corresponding means are $0.713 \pm 0.051$ for Uni-Macro, $0.702 \pm 0.148$ for SA-GCN, $0.665 \pm 0.097$ for Stage2-Only, $0.583 \pm 0.126$ for PerioGT, and $0.549 \pm 0.133$ for TransPolymer. For a direct comparison on identical records, we also evaluated all six models on the common 3461-record subset with nonidentical class predictions. FRPN achieved the highest pooled accuracy of 0.8229 (Table S9), showing that its advantage persists on records that are difficult for the model set to classify consistently.

The confusion matrices in **Figure 2**c further show that FRPN produces the most balanced predictions across the two classes. It achieves the largest numbers of correctly classified negative and positive samples, corresponding to the highest true-negative and true-positive counts among all models. The complete fold-wise confusion matrices for all six models are provided in Figure S2.

We next visualized the learned embedding space of evaluated models. The main t-SNE[44] visualization was generated with a perplexity of 50, while results obtained with other perplexity values are provided in Figure S3. In the probability-colored t-SNE as shown in **Figure 2**d and the label-colored t-SNE plots as shown in **Figure 2**e, FRPN shows regions with similar predicted probabilities and class labels, whereas the baseline models exhibit locally mixed patterns.

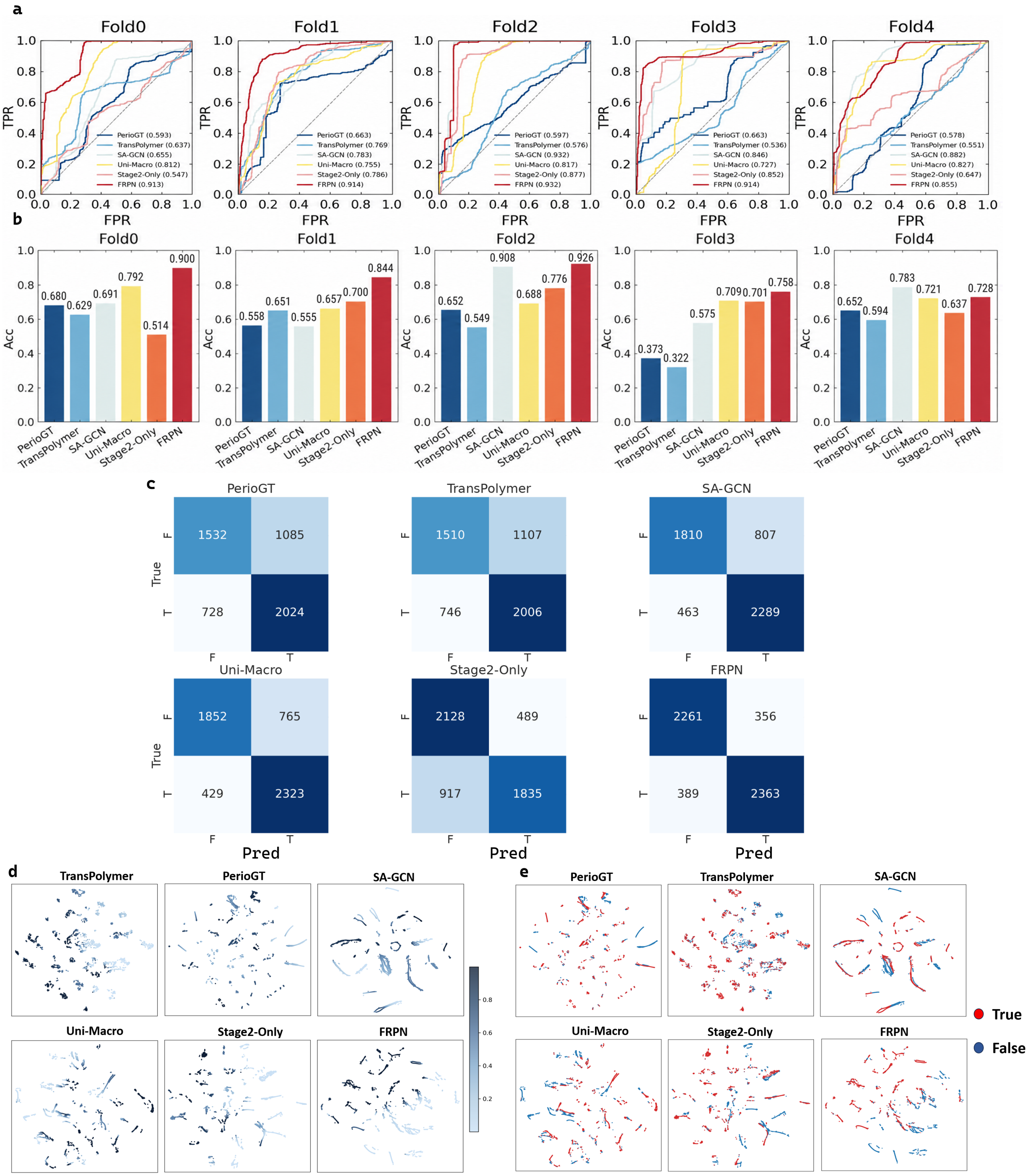

Figure 2. Out-of-fold BCDB classification performance and decision-level diagnostics. (a) Fold-wise ROC curves and AUC values. (b) Accuracy on model-specific disagreement subsets, defined for each model by inconsistent predictions among the other five models. (c) Confusion matrices aggregated across folds. (d) Embedding visualization colored by predicted probability. (e) Embedding visualization colored by ground-truth labels.

**Capacity and Representation Ablations.** To assess whether larger prediction heads recover the observed performance gain, we compare Uni-Macro-Big and Stage2-Big with their single-branch counterparts. FRPN-tiny provides a complementary control with reduced capacity and depth (Table S10). The corresponding model performances are reported in **Table 3.**

Enlarging the single-branch prediction heads yields modest improvements, while FRPN remains substantially better. FRPN-tiny also outperforms both enlarged single-branch controls despite its smaller capacity and depth. These results support the value of joint representation beyond the tested increases in prediction-head capacity.

**Table 3. Capacity ablation on the BCDB benchmark, with best values in bold and standard deviations across five folds.**

| Models | AUC | ACC | BACC | F1-score | MCC |
|---|---|---|---|---|---|
| Uni-Macro | 0.788±0.044 | 0.775±0.053 | 0.754±0.033 | 0.772±0.097 | 0.523±0.079 |
| Uni-Macro-Big | 0.802±0.017 | 0.783±0.031 | 0.763±0.013 | 0.788±0.078 | 0.547±0.025 |
| Stage2-Only | 0.741±0.141 | 0.737±0.120 | 0.724±0.111 | 0.695±0.161 | 0.461±0.210 |
| Stage2-Big | 0.749±0.179 | 0.787±0.091 | 0.752±0.119 | 0.710±0.227 | 0.539±0.226 |
| FRPN-tiny | 0.858±0.063 | 0.849±0.052 | 0.835±0.067 | 0.831±0.104 | 0.683±0.113 |
| FRPN | **0.906±0.029** | **0.864±0.054** | **0.853±0.073** | **0.848±0.108** | **0.717±0.120** |

We also compare FRPN with its variant FRPN Fingerprints (FRPN-FP) where monomer embeddings are derived from 2048-bit Morgan fingerprints[45] with radius 2, i.e., ECFP4 fingerprints. The results are summarized in **Table 4**. FRPN has a higher mean ROC–AUC than the fingerprint-based variant, while the other metrics are closely matched. This comparison indicates the advantage of learned monomer representations than fixed fingerprint embeddings. The ROC curves and OOF confusion matrices of FRPN-tiny, FRPN-FP, Uni-Macro-Big, and Stage2-Big are provided in Figure S4.

**Table 4. Representation ablation on the BCDB benchmark, with best values in bold and standard deviations across five folds.**

| Models | AUC | ACC | BACC | F1-score | MCC |
|---|---|---|---|---|---|
| FRPN-FP | 0.867±0.033 | 0.861±0.041 | 0.850±0.046 | **0.849±0.080** | 0.709±0.088 |
| FRPN | **0.906±0.029** | **0.864±0.054** | **0.853±0.073** | 0.848±0.108 | **0.717±0.120** |

We next examine symmetry breaking with and without volume-fraction information (Table S11). Anchor modulation improves the average results most clearly when volume fraction is omitted. This pattern suggests that the positional cues supplied by the anchor help distinguish block arrangements when a key compositional feature is absent.

**Performance evaluation on linear homopolymers.** FRPN has shown strong performance on the BCDB benchmark, where block copolymers carry non-trivial backbone-level information. However, whether this architecture remains beneficial for simple polymer systems, such as linear homopolymers, is not immediately clear. To examine this setting, we evaluate FRPN on a molecular-dynamics-derived linear homopolymer dataset reported by Yoshida et al.[46] We selected six representative properties, including density, radius of gyration (Rg), specific heat capacity (Cp), refractive index (RI), self-diffusion coefficient (D), and dielectric constant (EPS). Details of the dataset and standardization procedures are provided in Table S12.

**Table 5. Regression performance on the linear homopolymer benchmark, with best values in bold and standard deviations across five folds.**

| Label | Standardized RMSE | | | $R^2$ | | |
|---|---|---|---|---|---|---|
| | FRPN | Uni-Macro | Stage2-Only | FRPN | Uni-Macro | Stage2-Only |
| Density | 0.0966 ± 0.0220 | **0.0895 ± 0.0074** | 0.9124 ± 0.0993 | 0.9904 ± 0.0033 | **0.9918 ± 0.0006** | 0.1517 ± 0.0814 |
| Rg | 0.8699 ± 0.2136 | **0.7824 ± 0.2154** | 0.9906 ± 0.2597 | 0.2210 ± 0.1298 | **0.3775 ± 0.0910** | 0.0040 ± 0.0192 |
| Cp | 0.4811 ± 0.0903 | **0.4760 ± 0.1077** | 0.9745 ± 0.1160 | 0.7629 ± 0.0726 | **0.7664 ± 0.0845** | 0.0460 ± 0.0476 |
| D | 0.8594 ± 0.4779 | **0.8136 ± 0.4332** | 0.9750 ± 0.4206 | 0.2916 ± 0.1481 | **0.3572 ± 0.1445** | 0.0363 ± 0.0319 |
| EPS | 0.5899 ± 0.2273 | **0.4771 ± 0.1015** | 0.9947 ± 0.1895 | 0.6446 ± 0.1471 | **0.7609 ± 0.0573** | -0.0284 ± 0.0905 |
| RI | **0.1344 ± 0.0288** | 0.1427 ± 0.0331 | 0.9597 ± 0.0987 | **0.9813 ± 0.0068** | 0.9792 ± 0.0069 | 0.0736 ± 0.0486 |

**Figure 3.** Out-of-fold predicted versus true scatter plot on the linear homopolymer benchmark. (a) FRPN and Uni-Macro predictions. (b) Stage2-Only predictions.

The results are summarized in Table 5 and Figure 3. Uni-Macro has lower normalized RMSE and higher $R^2$ than FRPN on five of the six targets, while FRPN performs slightly better for RI. All samples share a linear homopolymer backbone, so topology and monomer ordering contribute little structural diversity. Uni-Macro's competitive performance highlights the predictive value of detailed monomer information in this setting. Stage2-Only has low $R^2$ across all six targets (Figure 3b), underscoring the value of those chemical features alongside chain structure. Together with the BCDB results, this comparison suggests that the benefit of explicit chain modeling depends on the structural diversity of the prediction task. RMSE is reported in standardized target space. Raw-unit RMSE values are provided in Table S13.

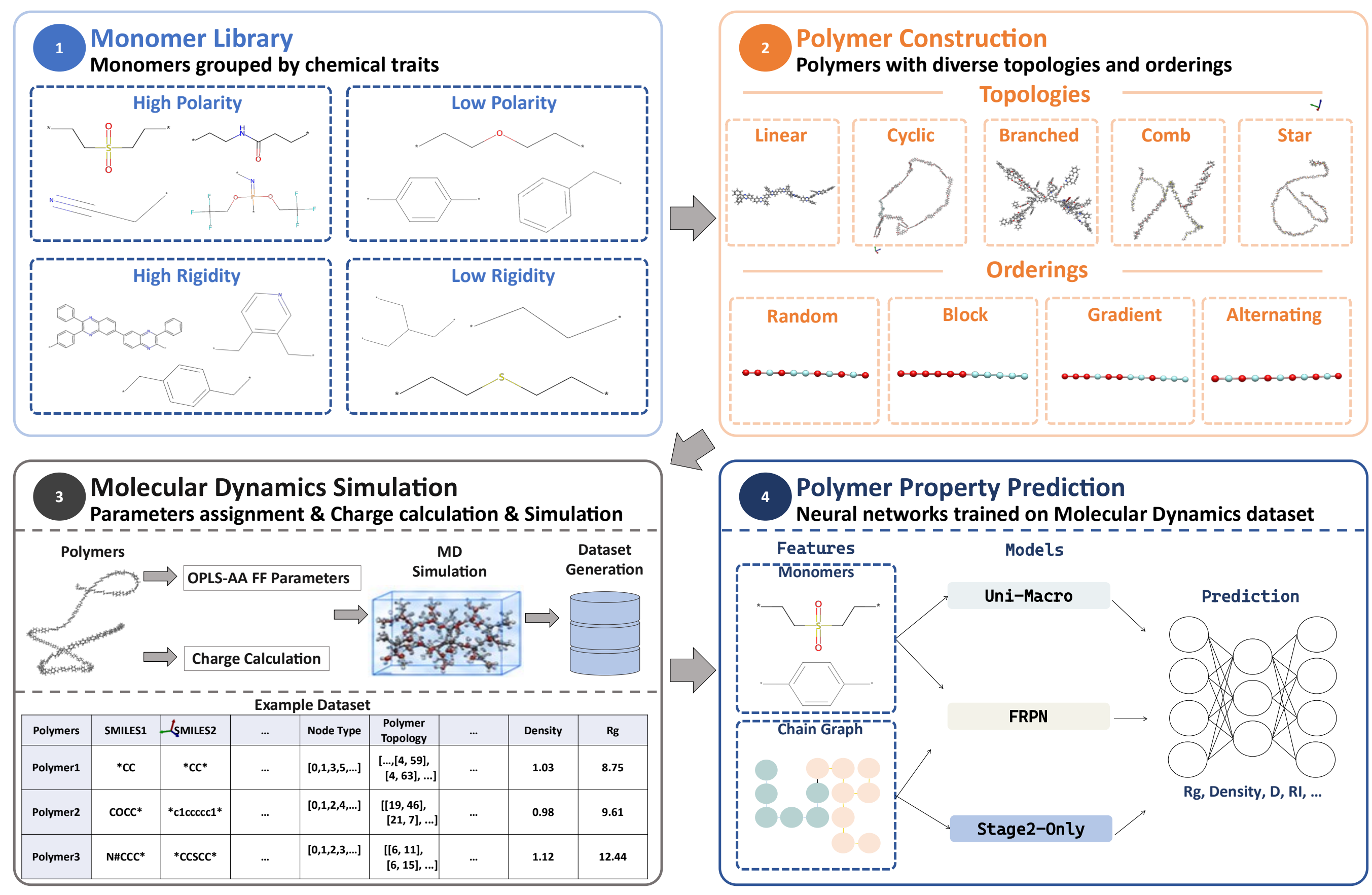

| Polymers | SMILES1 | SMILES2 | ... | Node Type | Polymer Topology | ... | Density | Rg |
|---|---|---|---|---|---|---|---|---|
| Polymer1 | *CC | *CC* | ... | [0,1,3,5,...] | [...,[4, 59], [4, 63], ...] | ... | 1.03 | 8.75 |
| Polymer2 | COCC* | *c1ccccc1* | ... | [0,1,2,4,...] | [[19, 46], [21, 7], ...] | ... | 0.98 | 9.61 |
| Polymer3 | N#CCC* | *CCSCC* | ... | [0,1,2,3,...] | [[6, 11], [6, 15], ...] | ... | 1.12 | 12.44 |

**Figure 4.** Construction of the structure-rich all-atom MD benchmark and downstream learning workflow. A monomer library spanning rigidity and polarity is assembled into polymers with diverse sequence orderings and chain topologies. All-atom MD simulations are performed under multiple temperature conditions to generate physical-property labels. The resulting benchmark provides explicit monomer chemistry, ordering, topology, and global-condition information for evaluating cross-scale polymer representation learning.

**MD benchmark for complex polymers.** To further evaluate FRPN beyond polymers with relatively simple structures, we constructed a molecular-dynamics-based benchmark spanning broader polymer properties and chain architectures. **Figure 4** summarizes the overall workflow used to construct the simulated polymer dataset and train property prediction models. We first built a monomer library that spans different chemical characteristics (the full monomer library is provided in Figure S5), such as polarity and rigidity, and then assembled polymers with diverse topologies and sequence orderings. Structures span linear, cyclic, branched, comb, and star graphs with 20–80 monomer instances. Construction and sequence-ordering definitions are in Note S1. MD simulations were then performed on LAMMPS[47] using the OPLS-AA[48] force field to obtain polymer property labels. To cover different thermal conditions, simulations were conducted in two temperature regimes, 300 K and 600 K. Both simulation settings followed a compression and decompression equilibration strategy adapted from the protocol of Larsen et al[49]. For the 300 K dataset, the system size was chosen following the RadonPy-style[50] polymer simulation setting, whereas the 600 K dataset used a larger 3×3×3 system construction. The detailed force-field assignment[48, 51], charge calculation[52] are provided in Note S2. Simulation settings and equilibration schedule[53, 54] are provided in Note S3.

Using this workflow, we obtained 1640 polymer structures spanning diverse monomer chemistries, topologies, and sequence orderings, with seven labels: density, Rg, D, EPS, nematic order (N. order), structure-factor peak height (S_q_peak), and RI. The labels define a learning task under the specified parameterization, system-construction, simulation, and extraction protocol within each temperature domain. Parameter completeness and charge consistency were audited (Note S2). The labels are finite-window property estimates, with convergence uncertainty described in Note S3. We compare the chemistry-focused Uni-Macro and chain-focused Stage2-Only controls with the joint FRPN and FRPN-FP models.

**Table 6. Model performance on the MD benchmark, with best values in bold and standard deviations across five folds.**

| Model | Density | Rg | D | EPS | N. Order | S_q_peak | RI |
|---|---|---|---|---|---|---|---|
| | | | **Metric: Standardized RMSE** | | | | |
| Uni-Macro | 0.1722±0.0410 | 0.2911±0.0238 | 0.2635±0.0227 | 0.3753±0.0473 | 0.8773±0.0489 | 0.8958±0.0708 | 0.0982±0.0262 |
| Stage2-Only | 0.0912±0.0098 | 0.2680±0.0160 | 0.2508±0.0210 | 0.3547±0.0705 | 0.8767±0.0482 | 0.8552±0.0673 | 0.0714±0.0131 |
| FRPN-FP | **0.0878±0.0109** | 0.2574±0.0172 | 0.2496±0.0221 | 0.3438±0.0643 | 0.8787±0.0512 | 0.8471±0.0648 | **0.0673±0.0133** |
| FRPN | 0.0879±0.0105 | **0.2509±0.0162** | **0.2460±0.0200** | **0.3297±0.0590** | **0.8712±0.0470** | **0.8245±0.0680** | 0.0684±0.0117 |
| | | | **Metric: $R^2$** | | | | |
| Uni-Macro | 0.9680±0.0125 | 0.9144±0.0157 | 0.9302±0.0104 | 0.8489±0.0374 | 0.2283±0.0189 | 0.1907±0.0284 | 0.9898±0.0056 |
| Stage2-Only | 0.9915±0.0018 | 0.9277±0.0101 | 0.9367±0.0097 | 0.8690±0.0330 | 0.2291±0.0250 | 0.2609±0.0529 | 0.9949±0.0015 |
| FRPN-FP | 0.9921±0.0019 | 0.9332±0.0104 | 0.9373±0.0103 | 0.8762±0.0307 | 0.2257±0.0284 | 0.2755±0.0392 | **0.9954±0.0014** |
| FRPN | **0.9922±0.0018** | **0.9365±0.0099** | **0.9392±0.0090** | **0.8866±0.0242** | **0.2389±0.0062** | **0.3146±0.0263** | 0.9953±0.0012 |

**Performance evaluation on complex polymer datasets**. We evaluate FRPN on the molecular-dynamics-based benchmark spanning complex polymer topologies and broad physical property targets. Details of the dataset, the splitting protocol, and standardization procedures are provided in Table S14. The results are summarized in Table 6. Overall, FRPN remains the best-performing model across most evaluated properties, with the lowest normalized RMSE on five of seven targets. FRPN-FP has slightly lower RMSE for density and refractive index. This pattern favors learned monomer representations overall within the tested joint architecture. RMSE on the MD benchmark is reported in standardized target space. Raw-unit RMSE values are provided in Table S15. Additional evaluation within each temperature subset shows that FRPN retains the best overall rank in both subsets (Table S16).

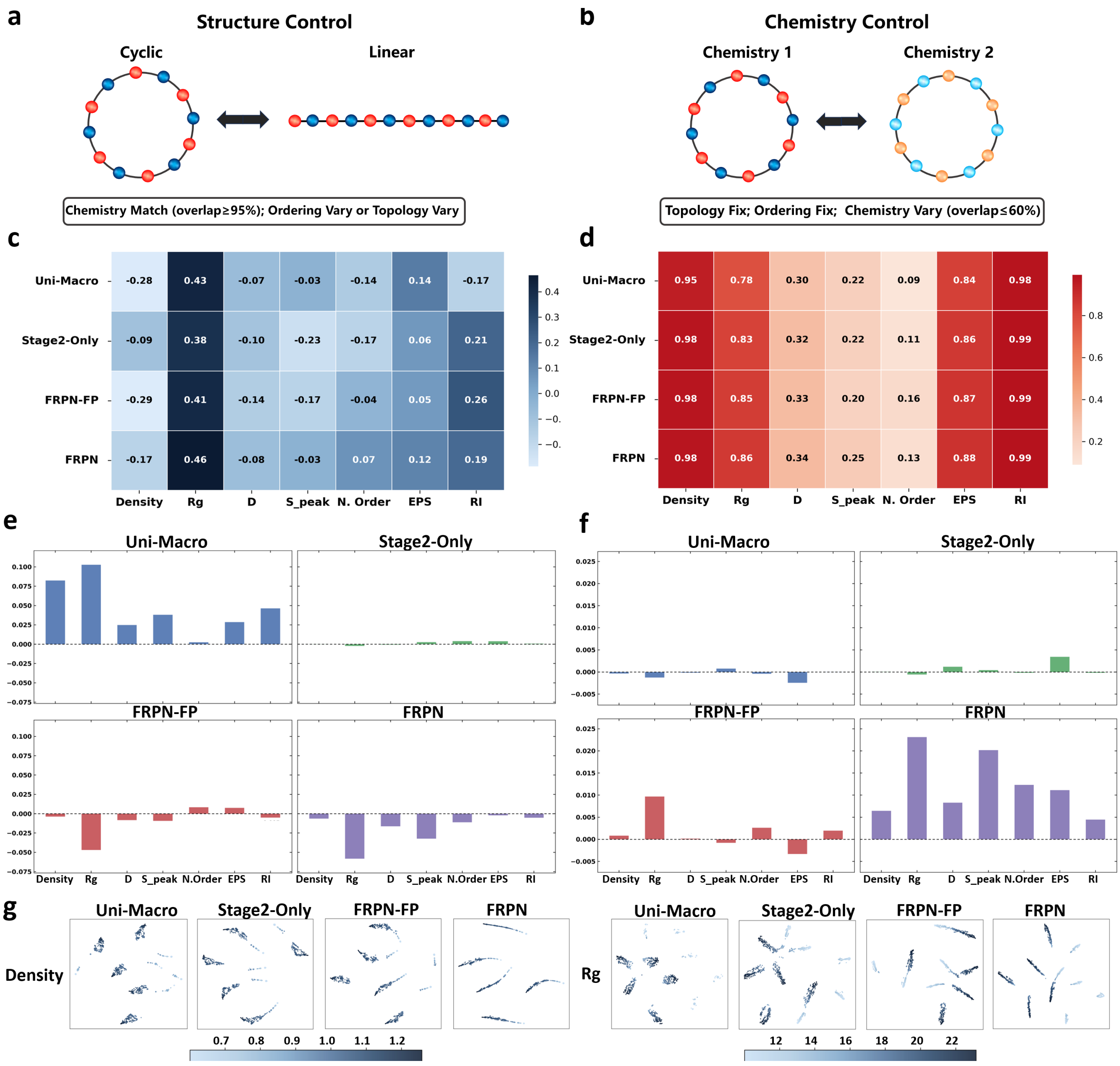

**Figure 5.** Representation diagnostics on the MD benchmark. (a) Schematic of the structure-control matched-pair analysis, where polymers have the same temperature and chain length, at least 95% monomer-composition overlap, and different topology, sequence ordering, or both. (b) Schematic of the chemistry-control matched-pair analysis, where polymers are matched by topology and ordering but differ in monomer chemistry. (c) Pearson correlation between z-scored true and predicted property differences for structure-control pairs. (d) The corresponding delta-correlation analysis for chemistry-control pairs. (e) Additive-null calibrated residual RMSE relative to the cross-fitted Uni-Macro plus Stage2-Only linear ensemble. (f) Leave-one-model-out residual stacking contribution, measured as the RMSE increase after removing each model from the four-model stack. (g) t-SNE projections of out-of-fold embeddings for density and Rg, colored by predicted property values.

**Representation diagnostics on the MD dataset.** The diverse structures and chemistries in the MD benchmark allow us to gain more insights from the model behaviors. We first perform a matched-pair analysis including a structure control setting as shown in Figure 5a, and a chemistry control setting as shown in Figure 5b. The branch inputs led us to expect Stage2-Only to track structural variation more consistently and Uni-Macro to track chemical variation more consistently.

The observed responses, however, do not follow this simple division. In the structure-control analysis, Uni-Macro has higher delta-correlations than Stage2-Only for five of seven targets (Table S17). Conversely, Stage2-Only has higher delta-correlations than Uni-Macro for six of seven chemistry-control targets (Table S18). These property-dependent patterns suggest that the single branches can exploit structure- or chemistry-related proxy signals from their available inputs. FRPN nevertheless performs best overall in the matched-pair comparisons, suggesting that cross-source proxy signals do not fully substitute for the benefits of explicitly providing both monomer chemistry and polymer structure within one architecture. These observations motivated two complementary follow-up analyses:

First, we compared residual prediction performance after accounting for the fitted additive main effects of chemistry group, topology, ordering, and temperature. Details of this method are provided in Note S5. As shown in Figure 5e, FRPN has lower residual RMSE than the cross-fitted Uni-Macro plus Stage2-Only linear ensemble for all seven targets (detailed metrics in Table S19). This suggests that the tested combination of separately learned predictions does not fully recover the benefit of joint modeling. One plausible explanation is that joint access allows chemical cues to be interpreted in the context of polymer structure before the final prediction.

Second, leave-one-model-out residual stacking examined whether a broader set of model outputs could replace FRPN's predictive contribution. As shown in Figure 5f, removing FRPN and refitting the stack with Uni-Macro, Stage2-Only, and FRPN-FP consistently increases residual RMSE for all seven targets (detailed metrics in Table S20). Importantly, FRPN-FP already combines fixed monomer fingerprints with the polymer graph. FRPN therefore provides a complementary contribution that is not fully supplied by the remaining predictions, and suggests the benefit of learning monomer semantics jointly in an end-to-end framework, rather than from fixed fingerprints.

Together, these diagnostics suggest that the design choice extends beyond which information sources are supplied to how monomer chemistry is learned and contextualized within the polymer chain. For the topology-rich MD benchmark, this provides a rationale for learning detailed monomer representations in conjunction with polymer-scale organization rather than treating the two sources only as separate predictors.

Finally, t-SNE projections of the OOF embeddings provide a view of property-related organization, with local gradients in predicted density and Rg (Figure 5g). Projections for the remaining targets are provided in Figure S6.

**Computational cost.** Table S21 reports inference latency and peak allocated GPU memory for the four principal models on an NVIDIA A800-SXM4-80GB. For FRPN, median batch-1 latency increases from 14.42 ms at 57 retained nodes to 103.52 ms at 1536 nodes, while peak memory increases from 535.3 to 2519.8 MiB. At 1536 nodes, FRPN-FP and Stage2-Only require 94.57 and 94.48 ms, respectively, compared with 8.98 ms for Uni-Macro. Table S21 gives the complete measurements and timing conditions.

## SUMMARY AND OUTLOOK

This work addresses a central representation problem in polymer informatics: how to learn polymer properties when the relevant structure is distributed across atomistic chemistry, monomer identity, and polymer topology. Uni-Macro-FRPN provides a full-resolution solution by coupling chemically resolved monomer encoding with polymer topology-aware contextualization. In this framework, monomers are not treated as isolated repeat-unit descriptors. They are embedded as chemically meaningful units whose roles are defined by their positions and connections within the polymer backbone.

The three benchmarks show how the benefit of full-chain reasoning depends on polymer structural diversity. On the BCDB benchmark, where block-copolymer organization is property-relevant, FRPN substantially improves over monomer-focused, chain-focused, and GNN-based chain-reasoning baselines. The capacity controls support the value of joint representation beyond larger prediction heads. On the linear homopolymer benchmark, detailed monomer representations remain competitive with the full-chain model, illustrating a simpler structural setting in which adding explicit chain organization yields limited benefit. On the structure-rich MD benchmark, where chemistry, ordering, topology, and physical conditions vary by design, FRPN achieves the strongest overall performance, with diagnostics supporting the value of learning monomer chemistry jointly with polymer-scale organization.

These findings, together with our previous Uni-Macro study, point to a more nuanced picture of polymer deep learning. Uni-Macro showed that monomer-centered representations can become performance-limited, while FRPN builds on this observation by introducing an explicit polymer-chain representation, suggesting that further progress may require not only strong monomer encoders, but also architectures that allow macromolecular organization to be learned directly.

Further work can extend both representation and evaluation. Sampling tacticity patterns and conformations would incorporate stereoregularity and instance-specific three-dimensional environments. More stringent chemical, structural, and source holdouts, together with matched cross-temperature studies under a harmonized simulation protocol, would assess extrapolation beyond the present comparisons. Extending the representation to interacting chains would then bring interchain packing and entanglement into the model.

## METHODS

**Stage-1 monomer encoder.** FRPN uses the previously reported Uni-Macro model as the stage-1 monomer encoder, including atomistic feature tokenization, pairwise structural features, and grouping-regulated attention. The output interface connects its learned monomer representations to the stage-2 polymer encoder as follows.

For a polymer sample containing $\boldsymbol{S}$ distinct monomer types and $\boldsymbol{N}$ atoms across these monomers, Uni-Macro receives a token sequence:

$$\boldsymbol{X}^{(1)} = [\boldsymbol{x}_{cls}^{(1)}, \boldsymbol{x}_{glob,1}^{(1)}, \dots, \boldsymbol{x}_{glob,G}^{(1)}, \boldsymbol{x}_{seg,1}^{(1)}, \dots, \boldsymbol{x}_{seg,S}^{(1)}, \boldsymbol{x}_{atom,1}^{(1)}, \dots, \boldsymbol{x}_{atom,N}^{(1)}]$$

Here, $\boldsymbol{x}_{cls}^{(1)}$ is the stage-1 summary token, $\boldsymbol{x}_{glob}^{(1)}$ is a global descriptor token for a feature such as temperature or pressure, $\boldsymbol{x}_{seg,s}^{(1)}$ is the segment token for the $\boldsymbol{s}$-th monomer type, and $\boldsymbol{x}_{atom,n}^{(1)}$ is the $\boldsymbol{n}$-th atom token. Uni-Macro maps this sequence to contextualized token representations:

$$\boldsymbol{H}^{(1)} = \left[\boldsymbol{h}_{\text{cls}}^{(1)}, \boldsymbol{h}_{\text{glob}}^{(1)}, \boldsymbol{h}_{\text{seg},1}^{(1)}, \dots, \boldsymbol{h}_{\text{seg},S}^{(1)}, \boldsymbol{h}_{\boldsymbol{atom},\boldsymbol{1}}^{(1)}, \dots, \boldsymbol{h}_{\boldsymbol{atom},\boldsymbol{N}}^{(1)}\right]$$

When Uni-Macro is used as a standalone baseline, the stage-1 summary representation is passed directly to the prediction head. When Uni-Macro is used inside FRPN, the segment-token representations $\boldsymbol{h}_{\boldsymbol{seg},s}^{(1)}$ are extracted as monomer embeddings for the stage-2 polymer encoder.

Stage-2 sequence. Node $\boldsymbol{i}$ has monomer type:

$$\boldsymbol{s}(\boldsymbol{i}) \in \{\boldsymbol{1}, \dots, \boldsymbol{S}\}, \qquad \boldsymbol{i} = \boldsymbol{1}, \dots, \boldsymbol{L}$$

For the full FRPN model, the initial chain-node token is initialized from the stage-1 monomer embedding:

$$\boldsymbol{x}_{\boldsymbol{i}}^{(2)} = \boldsymbol{h}_{\boldsymbol{seg},\boldsymbol{s}(\boldsymbol{i})}^{(1)}$$

For the Stage2-Only control model, the atomistic Uni-Macro encoder is bypassed and the node token is initialized from a learnable SMILES-identity embedding:

$$\boldsymbol{x}_{\boldsymbol{i}}^{(2)} = \boldsymbol{EMB}_{\boldsymbol{SMILES}}\big(\boldsymbol{s}(\boldsymbol{i})\big)$$

For the FRPN-FP control, the node token is initialized from a fixed molecular fingerprint followed by a small projection network:

$$\boldsymbol{x}_{\boldsymbol{i}}^{(2)} = \boldsymbol{W}_{\text{FP}}\, \boldsymbol{\phi}_{\boldsymbol{FP}}\big(\boldsymbol{s}(\boldsymbol{i})\big)$$

Where $\boldsymbol{\phi}_{\boldsymbol{FP}}$ denotes the molecular fingerprint of the corresponding monomer type. The complete stage-2 sequence is:

$$\boldsymbol{X}^{(2)} = \left[\boldsymbol{x}_{\boldsymbol{cls}}^{(2)}, \boldsymbol{x}_{\boldsymbol{glob},\boldsymbol{1}}^{(2)}, \dots, \boldsymbol{x}_{\boldsymbol{glob},\boldsymbol{G}}^{(2)}, \boldsymbol{x}_{\boldsymbol{monomer},\boldsymbol{1}}^{(2)}, \dots, \boldsymbol{x}_{\boldsymbol{monomer},\boldsymbol{L}}^{(2)}\right]$$

**Symmetry breaking for repeated chain nodes.** Largely repeated monomer instances can receive identical initial embeddings when they share the same monomer type. Without additional positional modulation, the stage-2 encoder may rely mainly on pooled monomer embeddings, rather than learning how identical monomers play different roles at different positions in the polymer structure. FRPN therefore introduces a topology-dependent symmetry-breaking modulation before stage-2 attention.

For a chain node $\boldsymbol{i}$, let $\boldsymbol{p_i}$ denote its normalized anchor-relative coordinate. The node token is rewritten as:

$$\widetilde{\boldsymbol{x}}_i^{(2)} = \boldsymbol{x}_i^{(2)} \odot [\boldsymbol{1} + \boldsymbol{MLP}(\boldsymbol{p_i})] + \boldsymbol{b_{topo}}$$

$\boldsymbol{MLP}$ denotes a multilayer perceptron, $\boldsymbol{b_{topo}}$ a learned topology bias, and $\odot$ element-wise multiplication. For BCDB, the anchor is the block junction. The signed distance $\boldsymbol{p_i}$ from node $\boldsymbol{i}$ is divided by the retained chain length plus $10^{-6}$. For MD, the graph center minimizes summed shortest-path distances. The model selects the first stored node when centers tie. Coordinates are normalized by maximum graph distance, bounded below by one. The graph-distance lookup is clipped at 10000 as a numerical safeguard. All distances in the evaluated graphs are below this bound. Cycles and multiple paths use their shortest length, and all 1640 evaluated MD graphs are connected. We evaluated graphs with multiple centers: predictions changed, but FRPN retained the best mean target rank (Table S22).

**BCDB graph construction and proportional rescaling.** DoPs determine the requested monomer counts in each block. For chains exceeding 1536 nodes, these counts are rescaled proportionally before token expansion. Integer counts are allocated from the proportional quotas while retaining each present monomer type and a total of 1536 monomer nodes. The allocated instances are permuted within blocks and joined linearly. Four special tokens are retained separately. Table S23 reports the within-block permutation analysis.

**Graph-distance-biased stage-2 attention.** The stage-2 encoder is a transformer-style graph attention encoder. Its pair bias is determined by the shortest-path distance $\boldsymbol{D_c}$ between chain nodes. For chain node $\boldsymbol{i}$ and chain node $\boldsymbol{j}$, let:

$$\boldsymbol{d_{ij}} = \boldsymbol{min}(\boldsymbol{D_c}(\boldsymbol{i}, \boldsymbol{j}), \boldsymbol{d_{max}})$$

Where $\boldsymbol{d_{max}}$ is a clipping distance, and $\boldsymbol{d_{i,j}}$ is then embedded as a pair representation:

$$\boldsymbol{b_{ij}} = \boldsymbol{EMB_{SPD}}(\boldsymbol{d_{ij}})$$

For attention head h in a stage-2 encoder layer, the attention logit from node $\boldsymbol{i}$ to node $\boldsymbol{j}$ is:

$$\boldsymbol{\alpha}_{ij}^{(h)} = \frac{\left(\left(\boldsymbol{q}_i^{(h)}\right)^{T} \boldsymbol{k}_j^{(h)}\right)}{\sqrt{\boldsymbol{d_h}}} + (\boldsymbol{w_h})^{T} \boldsymbol{b_{ij}} + \boldsymbol{M_{ij}}$$

Where $\boldsymbol{q}_i^{(h)}$ and $\boldsymbol{k}_j^{(h)}$ are the query and key vectors, $\boldsymbol{w_h}$ projects the pair representation to head $\boldsymbol{h}$, and $\boldsymbol{M_{ij}}$ is the attention mask. The normalized attention weights are obtained by applying softmax over valid tokens:

$$\boldsymbol{A}_{ij}^{(h)} = \boldsymbol{softmax}\left(\boldsymbol{\alpha}_{ij}^{(h)}\right)$$

The stage-2 encoder produces contextualized polymer representations $\boldsymbol{H}^{(2)}$, and the final prediction is made from the stage-2 summary representation $\boldsymbol{h}_{\text{cls}}^{(2)}$.

**Training protocol.** Each regression target used a separate single-output model with MSE loss on labeled records. BCDB used cross-entropy. ACC, BACC, F1, MCC, and confusion matrices used a fixed probability threshold of 0.5. FRPN and its controls used AdamW, cosine annealing, and gradient-norm clipping at 5. Regression checkpoints minimized validation RMSE, and classification checkpoints maximized validation accuracy. The same architecture and training configuration were used across folds, with no task-specific hyperparameter search (Table S3). Reported means and standard deviations summarize five folds, with one retained run per model and target per fold. Each held-out fold supplied both checkpoint selection and the reported score, so these scores may be optimistic. FRPN remained first in the available-fold mean validation-accuracy trajectory through epoch 90. Stopped folds were not carried forward (Figure S7).

**Molecular dynamics simulation.** All MD simulations were performed using LAMMPS with OPLS-AA-compatible bonded and nonbonded parameters. Detailed procedures for force-field assignment, charge calculation, and equilibration are provided in Notes S2 and S3.

The supervised targets comprise density, radius of gyration, self-diffusion coefficient, static structure-factor peak height, nematic order parameter, dielectric constant, and refractive index. Their calculation procedures are given below. Density was calculated as the time average of the simulation-cell density over the production window:

$$\boldsymbol{\rho} = \langle \boldsymbol{\rho}(\boldsymbol{t}) \rangle$$

The chain radius of gyration was calculated from per-chain gyration outputs and then averaged over chains and sampled frames:

$$\boldsymbol{R_g} = \left\langle \sqrt{\frac{\mathbf{1}}{\boldsymbol{M_c}} \sum_{i \in c} \boldsymbol{m}_i (|\boldsymbol{r}_i - \boldsymbol{r}_{COM,c}|^2)} \right\rangle_{c,t}$$

Here, $\boldsymbol{c}$ indexes polymer chains, $\boldsymbol{m_i}$ is the mass of atom $\mathbf{i}$, $\boldsymbol{M_c} = \sum_{i \in c} \boldsymbol{m}_i$, and $\boldsymbol{r}_{COM,c}$ is the chain center of mass of chain $\boldsymbol{c}$.

The self-diffusion coefficient was obtained from the mean-squared displacement using the Einstein relation:

$$\boldsymbol{D} = \frac{\mathbf{1}}{\mathbf{6}} \frac{\boldsymbol{d}}{\boldsymbol{dt}} \langle |\boldsymbol{\Delta r}(\boldsymbol{t})|^2 \rangle$$

In practice, the diffusion coefficient was estimated by fitting the slope of the late-time portion of the windowed MSD curve. The static structure factor was computed from wrapped atomic coordinates:

$$S(\boldsymbol{q}) = \frac{1}{N}\left|\sum_{j=1}^{N} exp(i\boldsymbol{q}\cdot \boldsymbol{r}_j)\right|^2$$

The reported S(q) peak label was defined as the maximum value on the scanned wave-vector grid after directional averaging, restricted to $\boldsymbol{q} \geq \mathbf{0.6}$ Å$^{-1}$.

The nematic order parameter was calculated from the largest eigenvalue of the orientational order tensor:

$$Q_{\alpha\beta} = \frac{1}{2}\langle 3u_\alpha u_\beta - \delta_{\alpha\beta}\rangle$$

$$S_{nem} = \lambda_{max}(Q)$$

Here, $\boldsymbol{u}$ is each molecule's principal covariance axis from unwrapped coordinates across topologies, $\boldsymbol{\delta_{\alpha\beta}}$ is the Kronecker delta, and $\boldsymbol{S_{nem}}$ is used here to avoid confusion with the static structure factor $\boldsymbol{S(q)}$.

The static dielectric constant was calculated from fluctuations of the total dipole moment of the simulation cell:

$$\varepsilon_r = 1 + \frac{\langle M^2\rangle - |\langle M\rangle|^2}{3\varepsilon_0 V k_B T}$$

Here, $\boldsymbol{M}$ is the total dipole moment, $\boldsymbol{V}$ is the cell volume, $\boldsymbol{T}$ is the temperature, $\boldsymbol{k_B}$ is the Boltzmann constant, and $\boldsymbol{\varepsilon_0}$ is the vacuum permittivity.

The refractive index was estimated using the Lorentz–Lorenz relation from the MD density and additive node-level molecular polarizabilities:

$$\frac{n^2 - 1}{n^2 + 2} = \frac{4\pi}{3}\frac{(N_A\rho)}{M}\alpha \times 10^{-24}$$

Here, $\boldsymbol{n}$ is the refractive index, $\boldsymbol{N_A}$ is Avogadro's constant, $\boldsymbol{M}$ is the molar mass of the repeat representation used for the additive polarizability, and $\boldsymbol{\alpha}$ is the corresponding molecular polarizability in Å$^3$. Node-level polarizabilities were calculated from H-capped node structures using finite-field Kohn–Sham DFT in PySCF,[55] with exchange-correlation functionals evaluated through Libxc.[56] Calculations were performed at the PBE0/def2-SVP level,[57-59] using Treutler–Ahlrichs radial grids and Becke partitioning for numerical integration.[60, 61]

## CODE AVAILABILITY

Code availability. Source code for data preprocessing, model implementation, training, and evaluation, together with the datasets and split identifiers used in this study, is available at github.com/PchGolden/Full-Resolution-Polymer-Nueral-Network. The GitHub package also provides the target-processing scripts and the final LAMMPS data files mapped to all 1640 records in the MD benchmark. Trained checkpoints and large cached evaluation assets will be archived on Zenodo, with checksums and download instructions maintained in the repository. The generalized force-field assignment workflow remains under development and will be released separately.

## SUPPORTING INFORMATION

Model architectures and training hyperparameters, dataset construction, simulation protocols, and target preprocessing, additional benchmark results and representation diagnostics are provided in the Supporting Information.

## AUTHOR INFORMATION

### Corresponding Author

Rui Zhang − South China Advanced Institute for Soft Matter Science and Technology, School of Emergent Soft Matter, South China University of Technology, Guangzhou 510640, China. Guangdong Provincial Key Laboratory of Functional and Intelligent Hybrid Materials and Devices, South China University of Technology, Guangzhou 510640, China. orcid.org/0000-0002-2099-9136

Email: rzhang1216@scut.edu.cn

### Authors

Jintao Wu − South China Advanced Institute for Soft Matter Science and Technology, School of Emergent Soft Matter, South China University of Technology, Guangzhou 510640, China. Guangdong Provincial Key Laboratory of Functional and Intelligent Hybrid Materials and Devices, South China University of Technology, Guangzhou 510640, China. orcid.org/0009-0001-3741-2143

Yiran Shan − South China Advanced Institute for Soft Matter Science and Technology, School of Emergent Soft Matter, South China University of Technology, Guangzhou 510640, China. Guangdong Provincial Key Laboratory of Functional and Intelligent Hybrid Materials and Devices, South China University of Technology, Guangzhou 510640, China. orcid.org/0009-0005-0483-3004

## ACKNOWLEDGMENT

This work was supported by the National Natural Science Foundation of China (No. 21973033) and the Natural Science Foundation of Guangdong Province (No. 2025A1515010995). The computations in this work were supported by High Performance Computing Platform of South China University of Technology.

## REFERENCES

(1) Guo, J., Schwaller, P. TANGO: direct optimization of constrained synthesizability for generative molecular design. *Nature Computational Science* **2026**, *6* (3), 260-270

(2) Lookman, T., Liu, Y., Gao, Z. Materials Informatics: Emergence to Autonomous Discovery in the Age of AI. *Advanced Materials* **2026**, *n/a* (n/a), e15941

(3) Wu, Y., Vriza, A., Ozgulbas, D., Vescovi, R., Zhou, J., Wang, Z., Hu, S., Zhang, Y., Yang, Q., Österholm, A. M., et al. Autonomous Synthesis and Inverse Design of Electrochromic Polymers with High Efficiency and Accuracy. *J Am Chem Soc* **2025**, *147* (48), 44101-44113

(4) Ge, W., De Silva, R., Fan, Y., Sisson, S. A., Stenzel, M. H. Machine Learning in Polymer Research. *Advanced Materials* **2025**, *37* (11), 2413695
(5) Wu, Y., Wang, C., Shen, X., Chen, Y., Wang, H., Xu, B., Zhu, Z., Chen, Y., Dai, W., Huang, Y., et al. Iterative discovery of potent polymeric antibiotics via multi-stage and multi-task learning against antimicrobial resistance. *Nature Communications* **2026**, *17* (1), 1878
(6) Van Krevelen, D. W., Te Nijenhuis, K. Chapter 1 - Polymer Properties. In *Properties of Polymers (Fourth Edition)*, Elsevier, 2009, pp 3-5.
(7) Lutz, J. F., Ouchi, M., Liu, D. R., Sawamoto, M. Sequence-controlled polymers. *Science* **2013**, *341* (6146), 1238149
(8) Wang, Z.-G. 50th Anniversary Perspective: Polymer Conformation—A Pedagogical Review. *Macromolecules* **2017**, *50* (23), 9073-9114
(9) Fetters, L. J., Lohse, D. J., Milner, S. T., Graessley, W. W. Packing Length Influence in Linear Polymer Melts on the Entanglement, Critical, and Reptation Molecular Weights. *Macromolecules* **1999**, *32* (20), 6847-6851
(10) Shen, C., Zhang, Y., Er, T. K. G., Han, F., Goto, A., Xia, K. Molecular Topological Deep Learning for Polymer Property Prediction. *ACS Nano* **2026**, *20* (1), 288-299
(11) Li, H., Zheng, H., Yue, T., Xie, Z., Yu, S., Zhou, J., Kapri, T., Wang, Y., Cao, Z., Zhao, H., et al. Machine learning-accelerated discovery of heat-resistant polysulfates for electrostatic energy storage. *Nature Energy* **2025**, *10* (1), 90-100
(12) Xu, C., Wang, Y., Barati Farimani, A. TransPolymer: a Transformer-based language model for polymer property predictions. *npj Computational Materials* **2023**, *9* (1), 64
(13) Kuenneth, C., Ramprasad, R. polyBERT: a chemical language model to enable fully machine-driven ultrafast polymer informatics. *Nat Commun* **2023**, *14* (1), 4099
(14) Qiu, H., Liu, L., Qiu, X., Dai, X., Ji, X., Sun, Z.-Y. PolyNC: a natural and chemical language model for the prediction of unified polymer properties. *Chemical Science* **2024**, *15* (2), 534-544
(15) Huang, Q., Li, Y., Zhu, L., Zhao, Q., Yu, W. Unified multimodal multidomain polymer representation for property prediction. *npj Computational Materials* **2025**, *11* (1), 153
(16) Chen, L., Kim, C., Batra, R., Lightstone, J. P., Wu, C., Li, Z., Deshmukh, A. A., Wang, Y., Tran, H. D., Vashishta, P., et al. Frequency-dependent dielectric constant prediction of polymers using machine learning. *npj Computational Materials* **2020**, *6* (1), 61
(17) Tamasi, M. J., Patel, R. A., Borca, C. H., Kosuri, S., Mugnier, H., Upadhya, R., Murthy, N. S., Webb, M. A., Gormley, A. J. Machine Learning on a Robotic Platform for the Design of Polymer–Protein Hybrids. *Advanced Materials* **2022**, *34* (30), 2201809
(18) Lin, T.-S., Coley, C. W., Mochigase, H., Beech, H. K., Wang, W., Wang, Z., Woods, E., Craig, S. L., Johnson, J. A., Kalow, J. A., et al. BigSMILES: A Structurally-Based Line Notation for Describing Macromolecules. *ACS Central Science* **2019**, *5* (9), 1523-1531
(19) Aldeghi, M., Coley, C. W. A graph representation of molecular ensembles for polymer property prediction. *Chem Sci* **2022**, *13* (35), 10486-10498
(20) Antoniuk, E. R., Li, P., Kailkhura, B., Hiszpanski, A. M. Representing Polymers as Periodic Graphs with Learned Descriptors for Accurate Polymer Property Predictions. *Journal of Chemical Information and Modeling* **2022**, *62* (22), 5435-5445
(21) Wu, Y., Wang, C., Shen, X., Zhang, T., Zhang, P., Ji, J. Periodicity-aware deep learning for polymers. *Nature Computational Science* **2025**, *5* (12), 1214-1226
(22) Wu, J., Zhang, R. Uni-Macro: A Flexible 3D Transformer for Multicomponent Molecular Systems with Insights into Polymer Deep Learning Paradigms. *Macromolecules 2026, 59 (13), 7367–7381.*
(23) Webb, M. A., Jackson, N. E., Gil, P. S., de Pablo, J. J. Targeted sequence design within the coarse-grained polymer genome. *Science Advances* **2020**, *6* (43), eabc6216
(24) Tao, L., Byrnes, J., Varshney, V., Li, Y. Machine learning strategies for the structure-property relationship of copolymers. *iScience 2022, 25 (7), 104585.*

(25) Hwang, W., Kwon, S., Lee, W. B., Kim, Y. Self-assembly prediction of architecture-controlled bottlebrush copolymers in solution using graph convolutional networks. *Soft Matter* **2024**, *20* (25), 4905-4915
(26) Kimmig, J., Köster, Y., Koswig, T., Raviswamy, P., Ganti, S. V. S., Zechel, S., Kuenneth, C., Schubert, U. S. Structure-Aware Machine Learning for Polymers: A Hierarchical Graph Network for Predicting Properties From Statistical Ensembles. *Macromolecular Rapid Communications* **2026**, *n/a* (n/a), e00671
(27) Han, M., Yokoo, T., Park, J., Oyaizu, K., Park, S. Deep learning prediction of ionic conductivity in polymer electrolytes using hierarchical polymer graphs. *Chemical Engineering Journal* **2025**, *521*, 166829
(28) Dwivedi, V. P., Rampášek, L., Galkin, M., Parviz, A., Wolf, G., Luu, A. T., Beaini, D. Long range graph benchmark. *Advances in Neural Information Processing Systems* **2022**, *35*, 22326-22340
(29) Topping, J., Di Giovanni, F., Chamberlain, B. P., Dong, X., Bronstein, M. M. Understanding over-squashing and bottlenecks on graphs via curvature. *arXiv preprint arXiv:2111.14522* **2021**,
(30) Caruso, A., Venturin, J., Giambagli, L., Rolando, E., El-Machachi, Z., Noé, F., Clementi, C. Extending the range of graph neural networks with global encodings. *Nature Communications* **2026**, *17* (1), 1855
(31) Alon, U., Yahav, E. ON THE BOTTLENECK OF GRAPH NEURAL NETWORKS AND ITS PRACTICAL IMPLICATIONS. In *9th International Conference on Learning Representations, ICLR 2021*, 2021.
(32) Muthukumar, M. 50th Anniversary Perspective: A Perspective on Polyelectrolyte Solutions. *Macromolecules* **2017**, *50* (24), 9528-9560
(33) Shirvanyants, D., Panyukov, S., Liao, Q., Rubinstein, M. Long-Range Correlations in a Polymer Chain Due to Its Connectivity. *Macromolecules* **2008**, *41* (4), 1475-1485
(34) Englebienne, P., Hilbers, P. A. J., Meijer, E. W., De Greef, T. F. A., Markvoort, A. J. Directional interactions in semiflexible single-chain polymer folding. *Soft Matter* **2012**, *8* (29), 7610-7616
(35) Garzon Otero, D., Akbari, O., Bilodeau, C. PepMNet: a hybrid deep learning model for predicting peptide properties using hierarchical graph representations. Molecular Systems Design & Engineering 2025, 10, 205–218.
(36) Zhang, J., Weng, X., Zhu, T., Liu, Y., Zhu, Z. Molecular-level protein semantic learning via structure-aware coarse-grained language modeling. Bioinformatics 2026, 42 (1), btaf654.
(37) Vaswani, A., Shazeer, N., Parmar, N., Uszkoreit, J., Jones, L., Gomez, A. N., Kaiser, L., Polosukhin, I. Attention is All you Need. In *Neural Information Processing Systems*, 2017.
(38) Rebello, N. J., Arora, A., Mochigase, H., Lin, T.-S., Shi, J., Audus, D. J., Muckley, E. S., Osmani, A., Olsen, B. D. The Block Copolymer Phase Behavior Database. *Journal of Chemical Information and Modeling* **2024**, *64* (16), 6464-6476
(39) Hanley, J. A., McNeil, B. J. The meaning and use of the area under a receiver operating characteristic (ROC) curve. *Radiology* **1982**, *143* (1), 29-36
(40) Bradley, A. P. The use of the area under the ROC curve in the evaluation of machine learning algorithms. *Pattern Recognition* **1997**, *30* (7), 1145-1159
(41) Brodersen, K. H., Ong, C. S., Stephan, K. E., Buhmann, J. M. The Balanced Accuracy and Its Posterior Distribution. In Proceedings of the 2010 20th International Conference on Pattern Recognition, 2010.
(42) Van Rijsbergen, C. J. Information retrieval. 2nd. newton, ma. USA: Butterworth-Heinemann: 1979.
(43) Matthews, B. W. Comparison of the predicted and observed secondary structure of T4 phage lysozyme. *Biochimica et Biophysica Acta (BBA) - Protein Structure* **1975**, *405* (2), 442-451
(44) Maaten, L. v. d., Hinton, G. E. Visualizing Data using t-SNE. *Journal of Machine Learning Research* **2008**, *9*, 2579-2605
(45) Rogers, D., Hahn, M. Extended-Connectivity Fingerprints. *Journal of Chemical Information and Modeling* **2010**, *50* (5), 742-754

(46) Yoshida, R., Hayashi, Y., Furuya, H., Hosoya, R., Kaneko, K., Sugisawa, H., Kaneko, Y., Takahashi, A., Noguchi, Y., Nanjo, S. Omics-scale polymer computational database transferable to real-world artificial intelligence applications. *arXiv preprint arXiv:2511.11626* **2025**,
(47) Plimpton, S. Fast Parallel Algorithms for Short-Range Molecular Dynamics. *Journal of Computational Physics* **1995**, *117* (1), 1-19
(48) Jorgensen, W. L., Maxwell, D. S., Tirado-Rives, J. Development and Testing of the OPLS All-Atom Force Field on Conformational Energetics and Properties of Organic Liquids. *Journal of the American Chemical Society* **1996**, *118* (45), 11225-11236
(49) Larsen, G. S., Lin, P., Hart, K. E., Colina, C. M. Molecular Simulations of PIM-1-like Polymers of Intrinsic Microporosity. *Macromolecules* **2011**, *44* (17), 6944-6951
(50) Hayashi, Y., Shiomi, J., Morikawa, J., Yoshida, R. RadonPy: automated physical property calculation using all-atom classical molecular dynamics simulations for polymer informatics. *npj Computational Materials* **2022**, *8* (1), 222
(51) Jorgensen, W. L., Tirado-Rives, J. Molecular modeling of organic and biomolecular systems using BOSS and MCPRO. *J Comput Chem* **2005**, *26* (16), 1689-1700
(52) Storer, J. W., Giesen, D. J., Cramer, C. J., Truhlar, D. G. Class IV charge models: A new semiempirical approach in quantum chemistry. *Journal of Computer-Aided Molecular Design* **1995**, *9* (1), 87-110
(53) MacKerell, A. D., Jr., Bashford, D., Bellott, M., Dunbrack, R. L., Jr., Evanseck, J. D., Field, M. J., Fischer, S., Gao, J., Guo, H., Ha, S., et al. All-Atom Empirical Potential for Molecular Modeling and Dynamics Studies of Proteins. *The Journal of Physical Chemistry B* **1998**, *102* (18), 3586-3616
(54) Ryckaert, J.-P., Ciccotti, G., Berendsen, H. J. C. Numerical integration of the cartesian equations of motion of a system with constraints: molecular dynamics of n-alkanes. *Journal of Computational Physics* **1977**, *23* (3), 327-341
(55) Sun, Q., Berkelbach, T. C., Blunt, N. S., Booth, G. H., Guo, S., Li, Z., Liu, J., McClain, J. D., Sayfutyarova, E. R., Sharma, S., et al. PySCF: the Python-based simulations of chemistry framework. *WIREs Computational Molecular Science* **2018**, *8* (1), e1340
(56) Lehtola, S., Steigemann, C., Oliveira, M. J. T., Marques, M. A. L. Recent developments in libxc — A comprehensive library of functionals for density functional theory. *SoftwareX* **2018**, *7*, 1-5
(57) Weigend, F., Ahlrichs, R. Balanced basis sets of split valence, triple zeta valence and quadruple zeta valence quality for H to Rn: Design and assessment of accuracy. *Physical Chemistry Chemical Physics* **2005**, *7* (18), 3297-3305
(58) Ernzerhof, M., Scuseria, G. E. Assessment of the Perdew–Burke–Ernzerhof exchange-correlation functional. *The Journal of Chemical Physics* **1999**, *110* (11), 5029-5036
(59) Adamo, C., Barone, V. Toward reliable density functional methods without adjustable parameters: The PBE0 model. *The Journal of Chemical Physics* **1999**, *110* (13), 6158-6170
(60) Treutler, O., Ahlrichs, R. Efficient molecular numerical integration schemes. *The Journal of Chemical Physics* **1995**, *102* (1), 346-354
(61) Becke, A. D. A multicenter numerical integration scheme for polyatomic molecules. *The Journal of Chemical Physics* **1988**, *88* (4), 2547-2553

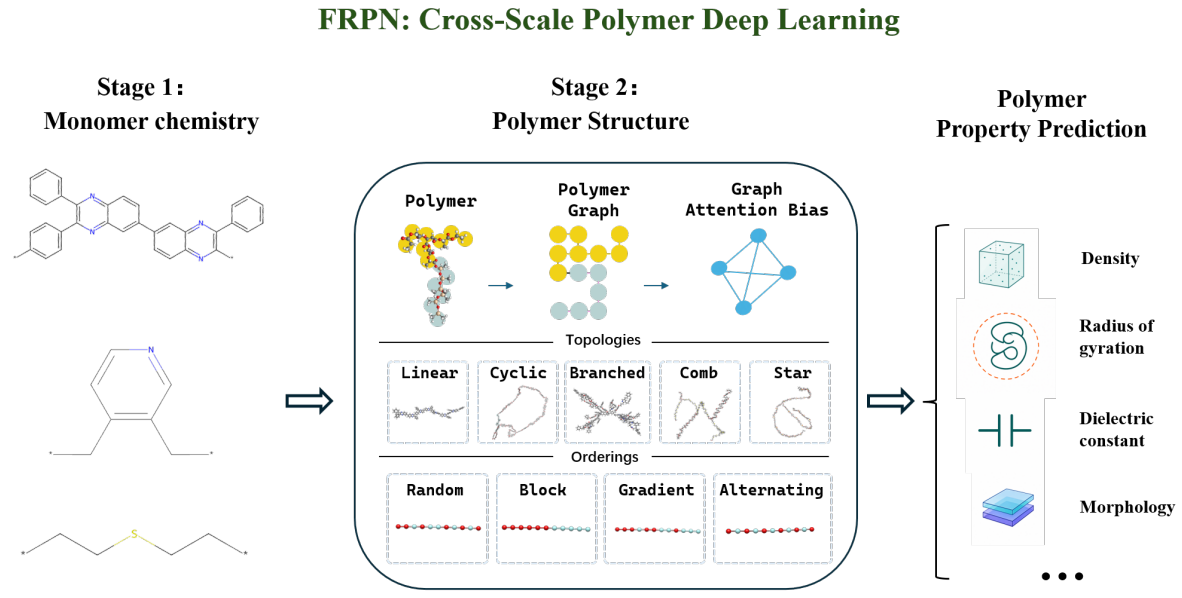